\documentclass[aps,showpacs,superscriptaddress,twocolumn,floats]{revtex4}
\usepackage{amssymb}
\usepackage{graphicx}
\begin{document}
\def\v#1{{\bf #1}}
\newcommand{\up}{\uparrow}
\newcommand{\dn}{\downarrow}
\newcommand{\la}{\langle}
\newcommand{\ra}{\rangle}
\newcommand{\lc}{\lowercase}
\newcommand{\La}{L\lc{a}M\lc{n}O$_3$~}
\newcommand{\LaBa}{L\lc{a}$_{1-x}$B\lc{a}$_x$M\lc{n}O$_3$~}
\title{Volume contraction at the Jahn-Teller transition of \La}
\author{T.~Maitra}
\affiliation{Max-Planck-Institute for the Physics of Complex Systems,
N\"othnitzer Str. 38, 01187 Dresden, Germany} 
\author{P.~Thalmeier}
\affiliation{Max-Planck-Institute for the Chemical Physics of Solids, 
N\"othnitzer Str.40, 01187 Dresden, Germany}
\author{T.~Chatterji}
\affiliation{Institute Laue-Langevin, BP 156, 38042 Grenoble Cedex 9, France}
\bibliographystyle{apsrev}
\begin{abstract}
We have studied the volume collapse of
LaMnO$_3$ at the Jahn- Teller (JT) transition temperature T$_{JT}$ =
750 K which has
recently been found in high temperature powder x- ray and neutron diffraction
experiments. We construct a model Hamiltonian involving
the pseudospin of Mn$^{3+}$ e$_g$ states, the staggered JT distortion and the
volume strain coordinate. We show that the anharmonic coupling between
these primary and secondary order parameters leads to the first order
JT phase transition associated with a comparatively large reduction of the unit
cell volume of $\Delta$V/V$\simeq$ 10$^{-2}$. We explain the
temperature dependence of JT distortions and volume strain and discuss
the volume change as function of the anharmonic coupling constant.
A continuous change to a second order transition as function of model
parameters is obtained. This behaviour is also observed under Ba doping.
\end{abstract} 

\maketitle

\section{Introduction}
The parent compound of colossal magnetoresistive
manganites, LaMnO$_3$ has drawn a lot of attention because of its
various states of spin and orbital order under variation of
temperature \cite{coey,dagotto,tokura}. This compound undergoes a structural
phase transition at T$_{JT}$ = 750 K associated with an
orbital order-disorder transition \cite{rodri}. It has been observed
that in the low temperature phase the orbital ordering is of C type
with alternate arrangement of $d_{3x^2-r^2}$ and $d_{3y^2-r^2}$
orbitals in the $ab$-plane while the planes are stacked along the
$c-$axis \cite{murakami}. On further lowering the temperature
LaMnO$_3$ undergoes a magnetic transition to an A-type
antiferromagnetic phase at the temperature T$_N\simeq$ 145 K where the spins
are aligned parallel to each other in the $ab-$plane and antiparallel
along the $c-$axis \cite{wollan,matsumoto}.

In a recent experiment T. Chatterji et al. \cite{tc1} have
investigated in detail the Jahn-Teller transition in LaMnO$_3$ using
high temperature x-ray and neutron diffraction on powder samples. They
observed that the unit cell volume of
LaMnO$_3$ decreases with increasing temperature in a narrow
temperature range below T$_{JT}$ and then undergoes a sudden collapse at
T$_{JT}$. It was argued that this striking volume
collapse is caused by the orbital order-disorder transition. In
the orbitally ordered phase the packing of MnO$_6$ octahedra needs
more space than in the disordered phase. The authors compared this
unusual phenomenon to the melting of ice which is accompanied by a similar
volume collapse. They also observed that with very small doping of Ba
the first order like transition becomes second order \cite{tc2}.
The above mentioned experimental observation of the volume
contraction at the Jahn-Teller transition temperature in LaMnO$_3$
has motivated us to study the effect from a microscopic model
for LaMnO$_3$. In addition to the usual
Jahn-Teller interaction terms we include the coupling between volume
strain and the Jahn-Teller distortions. This allows us to explain both
the observed volume contraction and the first order transitions in the
temperature dependence of Jahn-Teller distortions $Q_2$ and $Q_3$
within the same context.

\section{Pseudo spin model for the JT- transition in \La}
In \La the Mn$^{+3}$ ion in each MnO$_6$ octahedra is
in its $t_{2g}^3e_g^1$ state, the degenerate $e_g$ orbital is singly
occupied and hence JT-active. The local JT-distortions around each
Mn$^{+3}$ ion interact with each other cooperatively and give rise
to the observed orbital ordering. Several theoretical investigations
have been reported in the past to study the cooperative JT-phenomena
in 3d- compounds like MnF$_3$, KCuF$_3$, LaMnO$_3$ and
others. Among the first few attempts to study such systems is the
approach of Kanamori \cite{kana}. He started from a microscopic
Hamiltonian with the couplings of JT- electrons to uniform bulk
distortions as well as to all vibrational modes. In this model 
two possible ordered phases are considered, the `ferroorbital'
phase where the local distortions together with the orbitals of
JT-ions align in the same direction and the the
`antiferroorbital' case where they align in opposite directions
leading to a staggered order.
Kanamori's theory was successfully applied to some spinel-type
JT-crystals with ferroorbital order to explain the bulk tetragonal
distortion and the nature of the structural transition \cite{gg}. The theory 
of this type of ordering was further developed by Pytte \cite{pytte} for
more than one JT-ion in an unit cell. A review of JT phase
transitions was given by Sturge \cite{sturge}
In addition it was predicted that the structural phase transition for
the antiferro-orbitally ordered state should be of second order even
in the presence of higher order terms in the JT- energy and anharmonic
terms in the JT distortions \cite{kana}. 
But the observed phase transitions in most of these
antiferroorbitally ordered compounds are in fact first order like. 
Later on Kataoka \cite{kata} showed that for large higher order JT
interactions and if both ferro- and antiferro-
distortions are taken into account the transition may still be of
first order. 
We will now use a similar model, supplemented by the effect of the volume
strain, to the present case of \La. 
As in \cite{kana,pytte}, we start from a microscopic model Hamiltonian
with the usual first order JT-terms coupling the electronic states to the
$\Gamma_3$ (E$_g$) type JT-distortions. Since the fully symmetric
$\Gamma_1$ (A$_{1g}$) volume strain can always couple to the square of
the JT order parameter, irrespective whether it is of ferro- or antiferro-
type, one has to expect a corresponding third order anharmonic term in the
lattice energy. It has two consequences: Firstly it induces
a spontaneous volume strain  Q$_0$ below T$_{JT}$ as secondary order
parameter in addition to the primary staggered JT order parameter
Q$_s$. Secondly, if the coupling term is large enough compared to the
energy of the JT distortion, it can also lead to a first order
JT transition. Since in \La we indeed observe a large volume strain
below T$_{JT}$ and the transition is of first order, despite having a
pure staggered order parameter, we conclude that this term
which couples volume and JT- strains is more important than the higher
order terms in the JT energy itself introduced in \cite{kana,pytte,kata}. 
Our model Hamiltonian is then given by

\begin{eqnarray}
H&=&\frac{1}{2}C_T(Q_2^2+Q_3^2)-g_{0}
\sum_{i}(Q_{2}\sigma_{xi}+Q_{3}\sigma_{zi})\nonumber\\
&&+\frac{1}{2}C_BQ_0^2-\gamma_1Q_0(Q_2^2+Q_3^2)-\gamma_0Q_0^3
\label{Ham1}
\end{eqnarray}

Here the first term is the elastic energy of JT distortions with 
$C_{ij}=\Omega c_{ij}$ where $\Omega$ is the unit cell volume and 
for the ferro- distortive case $c_{ij}$ are the bulk elastic
constants. Thus C$_T$/$\Omega$ =
c$_{11}$-c$_{12}$ is the elastic constant of $\Gamma_3$ symmetry
distortions (Q$_2$,Q$_3$). In the ferro- distortive case they are
related to the cartesian
elastic strain components $e_{ii}$ by ($Q_2$,$Q_3$) = 1/$\sqrt{2}$((e$_{11}$
-e$_{22}$),1/$\sqrt{3}$(2e$_{33}$-e$_{22}$-e$_{11}$)). The second
term in Eq.~(\ref{Ham1}) represents the first order JT-coupling to the
distortions with $g_0$ being the coupling constant. The third term
gives the energy
due to volume change where the bulk modulus $C_B=1/3(C_{11}+2C_{12})$
and $Q_0$ is the volume strain given by
$Q_0=e_{11}+e_{22}+e_{33}$. Here we assume that $Q_0$(T) =
(V(T)-V$_0$(T))/V$_0$(T$_{JT}$) is the dimensionless volume change caused
by the JT ordering in addition to the already present backround volume
variation of V$_0$(T).
The coupling between volume strain and the JT-distortions is given by
the fourth term with a coupling strength $\gamma_1$. The last term
represents the anharmonic energy due to the volume change. Furthermore
$\sigma_x$ and $\sigma_z$ are pseudo spin Pauli matrices within the subspace of
e$_{1g}$ states. The experimental observations \cite{rodri}
show that the Mn-O bond lengths alternate between short and long in
the $xy$-plane and are the same along z-axis which suggests that pure staggered
orbital order is realised in \La without any admixture of
ferrodistortion . The pure staggered order is of C-type corresponding to
wave vector \v Q = ($\pi$,$\pi$,0).
To study this case we divide the lattice into two sublattices A and
B and the staggered order requires that distortions in A and B have
equal magnitude but opposite signs, i.e. $Q_{2A}=-Q_{2B}$
and $Q_{3A}=-Q_{3B}$). 
We write Q$_2$ and Q$_3$ in polar coordinates as Q$_2$ = Q$\sin{\theta}$
and Q$_3$ = Q$\cos{\theta}$. We then have Q$_A$ = -Q$_B$ = Q$_s$ or
Q$_s$ = 1/2(Q$_A$-Q$_B$). 

\begin{figure}
\includegraphics[clip=true,width=70mm]{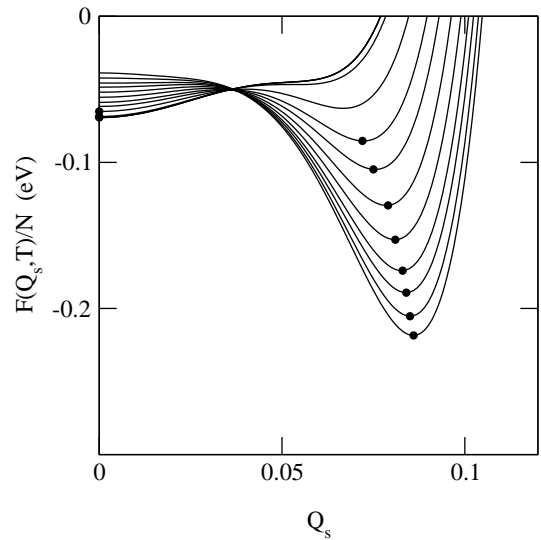}
\caption{The free energy $F(Q_s,T)$ per Mn$^{+3}$ ion is plotted against
the staggered JT-distortion $Q_s$ (the primary order parameter) for different
temperatures above and below the transition temperature. The value of the order
parameter for which the system has minimum free energy is shown by solid dots
for a sequence of temperatures starting at T/T$_{JT}$ = 0.6 (deepest
minimum on the right side) increasing in steps of 0.05 to a value of
T/T$_{JT}$ = 1.1. At T$_{JT}$ the order parameter Q$_s$ jumps to zero
from a finite value.}
\label{FREE}
\end{figure}

Thus in terms of Q$_s$ Eq.~(\ref{Ham1}) then becomes,

\begin{eqnarray}
H&=&\frac{1}{2}C_TQ_s^2-g_{0}Q_s
\sum_{i}(\sin{\theta} \sigma_{xi} +\cos{\theta}\sigma_{zi})\nonumber\\
&&+\frac{1}{2}C_BQ_0^2-\gamma_1Q_0Q_s^2
-\gamma_0Q_0^3
\label{Ham2}
\end{eqnarray}

Note that for the staggered case C$_T$ is not identical to the bulk
elastic constant and g$_0$ is different from the ferroorbital case,
however for simplicity we do not introduce new notations. The
Hamiltonian in Eq.~\ref{Ham2} is invariant under
rotations in the doubly degenerate e$_{1g}$ subspace, the
appropriate ground state orbitals $d_{3x^2-r^2}$ and $d_{3y^2-r^2}$ on
A- and B- sublattices corresponding to $\theta$ = 30$^\circ$ at low
temperatures are selected by higher order JT interactions
which we do not include explicitly, for the volume contraction effect
this is indeed  not necessary since the term proportional to
$\gamma_1$ is independent of the mixing angle $\theta$. The
Hamiltonian is then diagonalised and the free energy of the system is given by 

\begin{eqnarray}
F&=&-Nk_BT \log{Z(Q_s)}+NE_0\nonumber\\
Z(Q_s)&=&2\cosh{(g_0Q_s/k_BT)}\\
E_0&=&\frac{1}{2}C_TQ_s^2+\frac{1}{2}C_BQ_0^2
-\gamma_1Q_0Q_s^2-\gamma_0Q_0^3\nonumber
\end{eqnarray} 
 
In the following we will consider
the JT-distortion Q$_s$ as the primary order parameter of the
transition and the volume strain, Q$_0$, as the secondary order
parameter. Minimising the free energy with respect to Q$_0$ we arrive
at the relation 

\begin{equation}
Q_0=\lambda Q_s^2\qquad \mbox{for}\qquad 
\frac{\gamma_0\gamma_1}{C_B^2}Q_s^2 \ll 1
\label{SECOND}
\end{equation}

where we defined $\lambda=\gamma_1/C_B$. As shown below the condition
in Eq.~(\ref{SECOND}) is well fulfilled. Substituting
Eq.~(\ref{SECOND}) into the
expression of the free energy and minimising it with respect to Q$_s$
we get the mean-field equation for the staggered order parameter: 

\begin{eqnarray}
C_TQ_s -2\lambda^2C_BQ_s^3 -6\gamma_0\lambda^3Q_s^5 = 
g_0\tanh{(g_0Q_s/k_BT)}
\label{MEAN}
\end{eqnarray}

For the simple case $\lambda$ = 0 (no coupling of JT distortion and
volume strain) we have a second order transition with 

\begin{eqnarray}
k_BT_{JT}&=&\frac{g_0^2}{C_T}\qquad F(0)=-\frac{1}{2}k_BT_{JT}\nonumber\\
Q_s(0)&=&\frac{g_0}{C_T}\qquad
Q_0(0)=\lambda Q_s(0)^2=\frac{g_0^2\gamma_1}{C_BC_T^2}
\label{2ORDER}
\end{eqnarray}

For the general case $\lambda >$ 0 the solution for the transition temperature
T$_{JT}$ and the primary Q$_s$(T) and secondary (Eq.~(\ref{SECOND}))
Q$_0$(T) for T $<$ T$_{JT}$ has to be obtained numerically. The evolution
of the free energy and the
associated minimum described by the solution of Eq.~\ref{MEAN} is
shown in Fig.~\ref{FREE}. For the parameters chosen the first order
nature of the transition due to the $\gamma_1$ ($\lambda$) coupling term is
clearly obvious.

\begin{figure}
\includegraphics[clip=true,width=70mm]{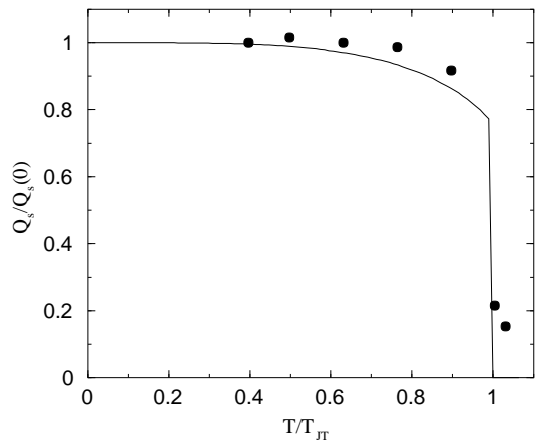}
\caption{Theoretical curve for the temperature dependence of
normalised JT-distortion (Q$_s$/Q$_{s}(0)$) (solid line) with parameters
C$_T$ = 85.0 GPa, $\lambda$ = 13.0, $\gamma_0$ = -400 GPa. The bulk
modulus C$_B$(T) is taken from \cite {darling} (see inset of
Fig.~\ref{Volstrain}). The
solid dots represents the experimental data \cite{tc1} for the
temperature dependence of Q$_s$.}
\label{JTstrain}
\end{figure}

\section{Numerical results and discussion} 
We now solve Eq.~\ref{MEAN} for the JT distortion numerically to obtain
the temperature
dependent order parameters . The energies per volume  are given in the unit of
GPa for convenience and the unit cell volume is taken from \cite{tc1}
to be V = 245.64\AA$^3$. Assuming that C$_B$ and C$_T$ are known from
measurements (this is only approximately true for C$_T$) we have three
fit parameters, the JT coupling g$_0$ and the anharmonic constants
$\gamma_0$ and $\gamma_1$, or equivalently $\lambda$=
$\gamma_1$/C$_B$. We fit these parameters consecutively. First the
JT-coupling strength g$_0$ = 1.898 GPa is
determined from the experimentally observed JT transition
temperature T$_{JT}$ = 750 K \cite{tc1} using the approximate relation
(Eq.~(\ref{2ORDER})) g$_0^2$/C$_T$ = k$_BT_{JT}$ valid for the second
order transition. Then the anharmonic coupling  parameters $\lambda$
(dimensionless) and $\gamma_0$ were determined to be 13.0 and
-400.0 GPa respectively by fitting the theoretical curves of Q$_s$(T)
and Q$_0$(T), including the jumps at T$_{JT}$, self-consistently to the
experimentally observed temperature dependence of of JT- distortion and
the unit cell volume \cite{tc1}.
 
\begin{figure}
\includegraphics[clip=true,width=70mm]{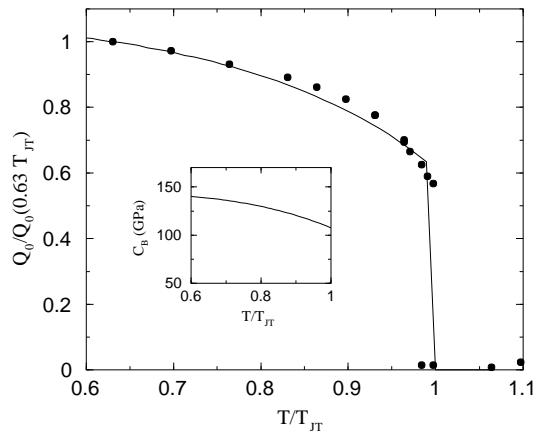}
\caption{Calculated temperature dependence of normalised volume
strain (Q$_0$/Q$_{0}(0.63T_{JT})$) (solid line) for the same set of
parameters as
in Fig.~\ref{JTstrain} along with the experimental data \cite{tc1} (solid dots)
for the temperature dependence of the normalised unit cell volume in
LaMnO$_3$. The temperature dependence of C$_B$ is extracted from
\cite{darling} (see inset).}
\label{Volstrain}
\end{figure}

Thereby we have also included the temperature
dependence of the bulk modulus C$_B$(T). It may be extracted from 
experimental results \cite{darling} obtained for Sr- doped \La by
properly rescaling T$_{JT}$ and fitting to the data in the temperature
range between JT- and magnetic transitions. The resulting C$_B$(T) is
shown in the inset of Fig.~\ref{Volstrain}. In Fig.~\ref{JTstrain} we show the
temperature dependence of the staggered JT-distortion Q$_s$
(normalised to its value at zero temperature)
from our calculation along with the corresponding experimental
points. The temperature dependence of the volume strain Q$_0$ (also
normalised by the value at zero temperature) calculated from our model
along with the corresponding experimental points for the unit cell
volume is presented in Fig. 3. With the parameter set used for
Figs.~\ref{JTstrain},\ref{Volstrain} and using the relations in
Eq.~\ref{2ORDER} we get an estimate 
$\frac{\gamma_0\gamma_1}{C_B^2}Q_s^2\simeq 0.9\cdot 10^{-2}$ for the
condition in Eq.~\ref{SECOND} in agreement with the assumption.
The parameters have been chosen to lead to the experimentally observed
first order phase transition. For illustration we show in Fig. 4(a)
that a reduction of the anharmonic coupling of primary (Q$_s$) and
secondary (Q$_0$) order parameters leads to a change from first to
second order phase transition as witnessed by the vanishing jump in
the order parameters and the associated vanishing of the latent heat.

\begin{figure}
\includegraphics[width=70mm]{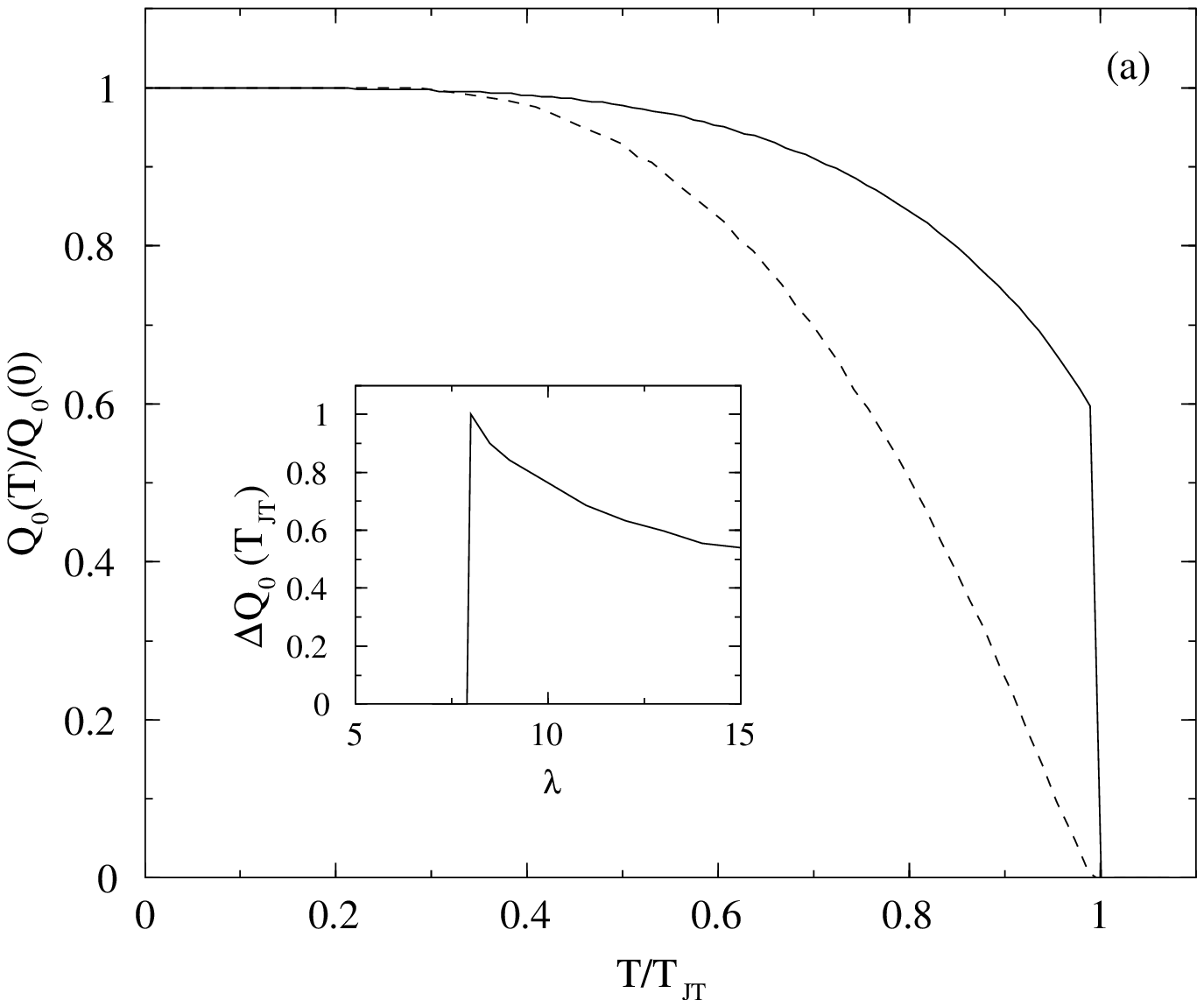}\\
\includegraphics[clip=true,width=70mm]{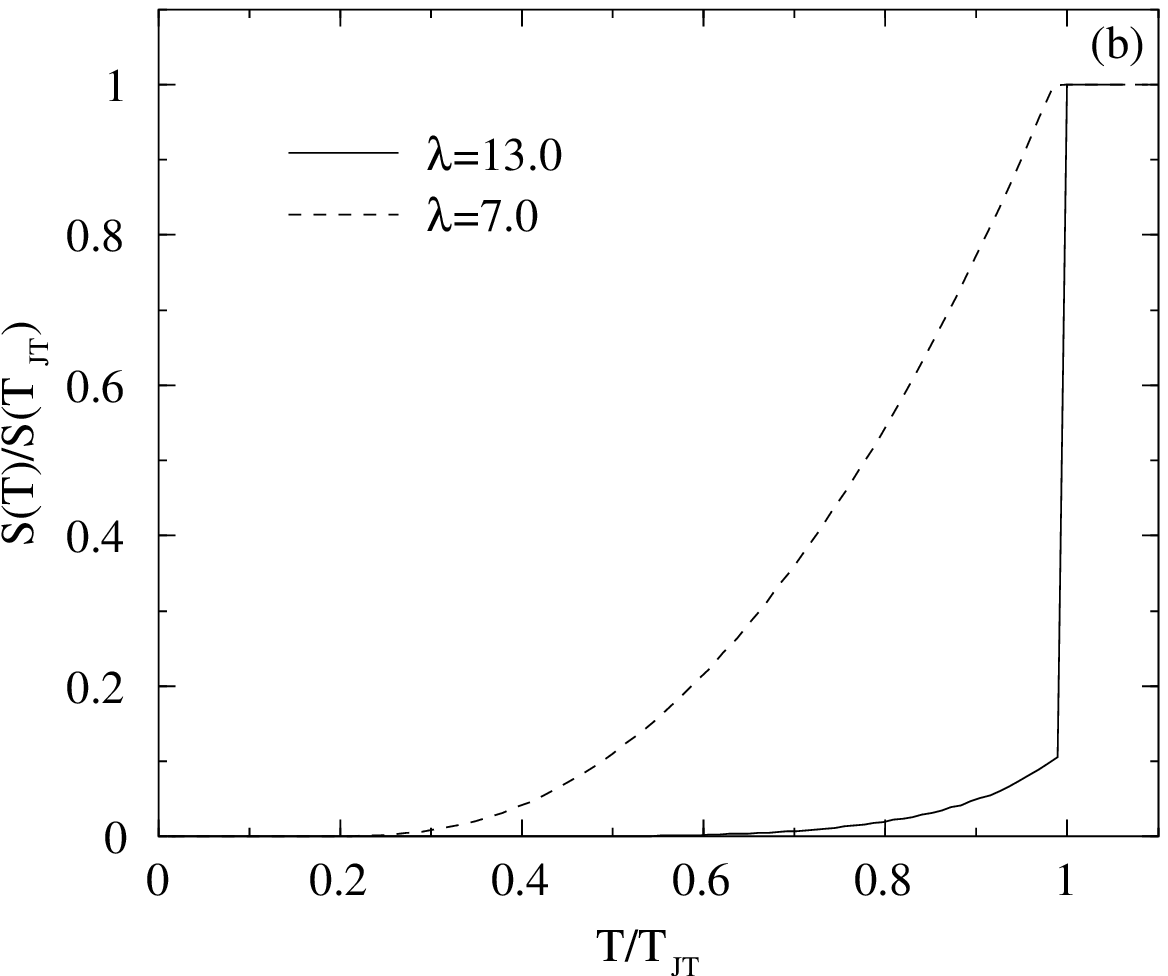}
\caption{The temperature dependence of $Q_0$ normalised by its value at 
$T=0$ is shown for two different values of $\lambda$. The curves for 
$\lambda=13$ (solid line) and $\lambda=7$ (dashed line) show first and 
second order transitions respectively. The inset in (a) gives the dependence
of the jump in $Q_0$ at the transition temperature on $\lambda$. In (b) we
show the corresponding curves for the temperature dependence of
entropy normalised at T$_{JT}$ for the same set of values of $\lambda$.}
\label{FS}
\end{figure}

In the inset of Fig. 4(a) we show that the volume jump actually
first increases when $\lambda$ is reduced and then suddenly drops to zero. 
This transition from first to second order is indeed observed for Ba-
doped \LaBa \cite{tc2} where the volume jump vanishes already for x = 0.025.
In Fig. 4(b) we show the temperature dependence of the entropy changing
discontinuously (continuously) with temperature at the transition for first 
(second) order transition. 

In this work we have given an explanation to the
experimentally observed first order JT- transition in
LaMnO$_3$. The first order transition is due to a strong coupling of
primary JT-distortion and secondary volume strain order
parameters. This result differs from those of previous models
\cite{kana} which did  not include the
coupling to the volume strain but rather consider higher order terms
in the JT energy. There the temperature dependence of JT-distortions
is of second order type in the pure anti-ferro orbital
order. Therefore the earlier models cannot be directly applied to
the present case of \La where one has a first order transition despite
having a pure staggered order parameter.
>From our calculation we see that the term which couples volume strain
and JT-distortion in the model Hamiltonian is also responsible for the sharp
collapse of  the bulk volume. In the absence of such coupling
\cite{kana,pytte}, the JT-distortions are volume conserving. 

We have not considered the spin degrees of freedom  in our model
calculation as has been discussed in \cite{kugel} and \cite{okamoto}
where it is argued that spin and orbital ordering
can interfere with each other through the orbital dependent
superexchange interaction. Since the magnetic transition temperature
(T$_N\simeq$ 145 K) in \La is much lower than the orbital ordering
temperature at T$_{JT}\simeq$ 750 K, we neglected the effect of
spin correlations on the orbital order.

\section{Conclusion} 

The pronounced volume collapse in \La at the JT-transition temperature
has been studied within a JT pseudospin model. Such models have
been used previously to describe structural phase transitions with a
JT distorion as order parameter in spinel- type crystals
\cite{kana,pytte}. To explain the additional volume
contraction as secondary order parameter at the orbital
order-disorder transition we have supplemented the model with an
anharmonic 
interaction between bulk volume strain and the JT-distortion. 
We have studied the evolution from second to first order phase
transition as function of the coupling strength and give a realistic
parameter set valid for \La from comparison of the observed T$_{JT}$
and the associated jump in the JT distortion and the volume collapse.
We also have shown that under suitable change of coupling parameters
the transition changes from first to second order which may simulate
the observed behaviour under small Ba- doping of the crystals.

\end{document}